\documentclass[12pt,a4paper]{article}
\usepackage{amsfonts,amsmath,amsthm}
\usepackage{graphicx}
\usepackage{epsfig}
\usepackage{slashed}
\usepackage{amssymb}
\usepackage{color}

\newtheorem{thm}{Theorem}[section]
\newtheorem{cor}[thm]{Corollary}

\newcommand{\be}{\begin{equation}}
\newcommand{\ee}{\end{equation}}
\newcommand{\bea}{\begin{eqnarray}}
\newcommand{\eea}{\end{eqnarray}}
\newcommand{\beq}{\begin{eqnarray}}
\newcommand{\eeq}{\end{eqnarray}}
\newcommand{\fin}{\begin{flushright}$\blacksquare$\end{flushright}}

\newcommand{\nn}{\nonumber}
\newcommand{\pa}{\partial}

\def\sD{\slashed{D}}

\def\({\left(}
\def\){\right)}
\def\[{\left[}
\def\]{\right]}
\def\<{\left<}
\def\>{\right>}
\def\a{\alpha}

\def\g{\gamma}

\begin{document}


\title{Zeroes of combinations of Bessel functions and mean charge of graphene nanodots}


\author{{C. G. Beneventano$^\dag$, I. V. Fialkovsky$^\ddag$, E. M. Santangelo$^\dag$}\\{\it $^\dag$ Departamento de F\'isica, Universidad Nacional de La Plata}\\
{\it Instituto de F\'isica de La Plata,  CONICET--Universidad Nacional de La Plata,}\\
{\it C.C.67, 1900 La Plata, Argentina}\\
{\it $^\ddag$ CMCC-Universidade Federal do ABC, Santo Andr\'e, S.P., Brazil}\\
{and \it Department of Theoretical Physics, Saint-Petersburg State University, }\\
{\it St. Petersburg 198504, Russia}
}

\maketitle





\maketitle

\begin{abstract}

We establish some properties of the zeroes of sums and differences of contiguous Bessel functions of the first kind. As a byproduct, we also prove that the zeroes of the derivatives of Bessel functions of the first kind of different orders are interlaced the same way as the zeroes of Bessel functions themselves. As a physical motivation, we consider gated graphene nanodots subject to Berry-Mondragon boundary conditions. We determine the allowed energy levels and calculate the mean charge at zero temperature. We discuss in detail its dependence on the gate (chemical) potential. The effect of temperature is also commented.

\end{abstract}
Bessel functions, graphene, quantum nanodots, circular billiards

\section{Introduction}
\label{sec-introduction}

Bessel functions are among the most ubiquitous special functions of mathematical physics. They frequently appear in the study of problems with both cylindrical and spherical symmetries. In particular, the properties of their zeroes, of the zeroes of their derivatives and of combinations of different Bessel functions and/or their derivatives frequently appear when solving boundary value problems of physical interest \cite{courant}.

Some properties of such zeroes have been known since long ago \cite{watson}. Others, have been proved more recently. To give only a few examples, in \cite{ifantis1} some bounds for the first positive zero of a combination (involving the zero itself) of a Bessel function and its derivative were obtained. Reference \cite{ifantis2} presents a study of zeroes of other combinations of Bessel functions of the first kind and their derivatives, including the zeroes of the second derivative as a particular case. Several results concerning zeroes of Bessel functions are presented in \cite{elbert}. The problem of the eventual coincidence of zeroes of Bessel functions of different orders, when such orders differ by an amount which is not a positive integer is addressed in \cite{petro}. Studies of the interlacing of zeroes of Bessel functions of first and second kinds and of their derivatives are contained in \cite{zou,palmai,palmai2}. Finally, \cite{esposito} presents an analysis of the zeroes of other combinations of Bessel functions of the first kind and their derivatives (equivalently, zeroes of combinations of Bessel functions of consecutive orders which again depend on the zeroes themselves), a problem arising in connection with the study of Euclidean quantum gravity on the four-dimensional ball.

In this paper, we study the relative positions of the zeroes of sums and differences of Bessel functions of the first kind of consecutive (real and non-negative) orders and their interlacing. As a byproduct, we prove that the zeroes of the derivatives of such Bessel functions are interlaced in a way identical to the well-known interlacing of the zeroes of the corresponding Bessel functions.


The consideration of the linear combinations of Bessel functions just described is motivated here by the investigation of the spectrum and the mean charge of a graphene circular nanodot, and the dependence of the charge on the externally applied gate potential. Within  the continuum limit (Dirac theory) such a nanodot can be described by a graphene disk with Berry-Mondragon \cite{BerryM} (or MIT bag \cite{MITbag}) boundary conditions (BCs) imposed on the Dirac fermions at the boundary. Berry-Mondragon BCs are one in a family of boundary conditions leading to a self-adjoint Dirac Hamiltonian \cite{Ben-Sant}.

The exceptional properties of graphene, a single atomic layer of carbon atoms, are widely known. So, we will not give a detailed list of them here. The interested reader is referred, for instance, to \cite{grarev,RMP,RevDasSarma11,KatsBook}. The fact that graphene is described, in the continuum limit, by a massless Dirac equation was demonstrated theoretically in \cite{Semenoff84,DiVincenzo84}, and experimentally confirmed more than twenty years later \cite{1}. Since then, many of the predictions of a ``relativistic'' massless Dirac theory and of field-theoretic methods as applied to this new material have been confirmed \cite{QFTGra,Maria,nos}. Among such predictions are the absence of a gap in infinite samples of graphene and a non--vanishing minimal conductivity which, together, constitute a major obstacle for further application of this two-dimensional material to controllable electronic devices. This fact triggered a huge number of studies aimed at opening an energy gap in graphene. One of the possible ways of achieving this goal is the use of samples of finite size, like nanoribbons and nanodots. An overview of the state of the art in this field, along with further literature, can be found in several recently published review articles on the transport and electronic properties of graphene nanostructures, including nanoribbons and nanodots \cite{RMP,RevDasSarma11,RevDubois09, RevStampfer2011,RevSnook2011}.

In a recent article, we studied the mean charge density and the longitudinal conductivity of disordered graphene nanoribbons when Berry-Mondragon boundary conditions are imposed at their edges \cite{bfsv} and compared our results with the available experimental data. In this paper, we use the methods of quantum field theory (QFT) to evaluate the mean charge in a graphene nanodot with the geometry of a disk and to reveal its dependence on the applied gate voltage. The model we use proved to be applicable to the description of graphene nanodots  of intermediate size  \cite{Ponomarenko2008}. Our present calculation shows that the mean charge presents jumps of equal height each time the gate voltage reaches the value of an allowed eigenenergy. Since such eigenenergies are determined by the zeroes of sums and differences of Bessel functions of the first kind of consecutive integer orders, a precise knowledge of the positions and interlacing properties of such zeroes is useful in order to predict the number of jumps in a given range of the gate voltage.

The outlay of the manuscript is the following.
We start by introducing the notation and formulating the physical problem, which requires the investigation of the Bessel functions: in Section \ref{sec-conventions} we present the conventions adopted and determine the spectrum of graphene nanodisks.
We continue in Section \ref{sec-dif} by stating and proving the Theorem \ref{teorema1} concerning the positions of the zeroes of differences of Bessel functions of the first kind of orders differing by one. Its corollary gives the precise interlacing of such zeroes. In section \ref{sec-sum} we prove, also for nonnegative orders, a theorem (Theorem \ref{teorema2}) concerning the ordering of zeroes of sums of the same Bessel functions. Their interlacing with the zeroes studied in the previous section and with the zeroes of Bessel functions and of their derivatives result as corollaries. A new proof of the interlacing of the zeroes of the derivatives of Bessel functions, which is identical to that of the zeroes of Bessel functions, is also obtained.  Although our physical problem requires only the consideration of integer orders of Bessel functions, we get more general results, holding for real nonnegative orders. Finally, in Section \ref{sec-charge} we present the calculation of the mean charge of Berry-Mondragon graphene nanodisks through field-theoretical methods \cite{chodos}, while some concluding remarks and comments appear in Section \ref{sec-remarks}.

\section{Graphene nanodisks. Conventions and energy spectrum}\label{sec-conventions}

In the continuum description of charge carriers in graphene, which has been found to be quite accurate \cite{grarev}, the behaviour of the wave function is governed by the Dirac equation \cite{Semenoff84,DiVincenzo84}. For a single Dirac cone (valley) its covariant form can be written as
\begin{equation}
  \sD \psi(x)=0\,,
\label{diracop}\end{equation}
where  ${\psi}=(\psi_1,\psi_2)^T$  is a two component spinor,
\begin{equation}\nn
    \sD=i\tilde \gamma^\mu \partial_\mu, \qquad
    (\tilde\gamma)\equiv (\gamma^0,v_F\gamma^1,v_F\gamma^2),
\end{equation}
and $\g^{0,1,2}$ are $2\times2$ Dirac (gamma) matrices in either of the two nonequivalent representations of the Clifford algebra in $2+1$ dimensions. Here and below we work in natural units, $\hbar=c=1$. In these units, the Fermi velocity $v_F\approx 1/300$.

We choose, in one of the two valleys, the following representation of the Dirac matrices in Minkowski space-time (with metric $(+,-,-)$)
\begin{equation}\nn
    \gamma^0=
    \left( \begin{array}{cc} 1 & 0 \\ 0 &-1\end{array} \right),\quad
    \gamma^1=
    \left( \begin{array}{cc} 0 & -1 \\1 & 0 \end{array} \right),\quad
    \gamma^2=
    \left( \begin{array}{cc} 0 & i \\ i & 0 \end{array} \right).
\end{equation}
Since we consider the parity--even configuration, the contributions from different valleys and for different spins just sum up; thus, to obtain results for a real graphene sample, one has to introduce the degeneracy factor, $N = 4$.

In order to treat the problem at hand, i.e., the problem of a graphene disk of radius $R$, polar coordinates $(r, \theta)$ are better suited. In such coordinates, the Hamiltonian, $H$, corresponding to eq. (\ref{diracop}) in the absence of external fields, is given by
\beq
\frac{1}{v_F}H=i{\gamma}^0{\gamma}^r \frac{\partial}{\partial r}+\frac{i}{r}{\gamma}^0{\gamma}^{\theta}\frac{\partial}{{\partial}{\theta}}\,,
\label{hamilton}\eeq
with
\begin{equation}\nn
    \gamma^r=
    \left( \begin{array}{cc} 0 & e^{-i\theta}\\-e^{i\theta} & 0 \end{array} \right),\quad
    \gamma^{\theta}=
    \left( \begin{array}{cc} 0 & -i e^{-i\theta}\\ -ie^{i\theta} & 0 \end{array} \right).
\end{equation}

The Berry--Modragon BCs, which we impose at the boundaries of the nanodot, were developed by considering fermions localized to a compact region due to a `locking' infinite mass potential \cite{BerryM}. They are the $2+1$ analogue of the so-called MIT bag BCs, introduced to model confinement in Quantum Chromodynamics in $3+1$ dimensions \cite{MITbag}. They imply zero current flux in the direction perpendicular to the boundary. Written in a $\gamma$-representation-independent way, they read
\begin{equation}
	\left.\frac{1+i{\gamma}^{\mu}n_{\mu}}{2}\psi \right|_{B}=0 \,,
\end{equation}
where $B$ is the boundary of the region to be considered and $n_{\mu}$ is the $2+1$--dimensional external normal vector at the boundary.
In the present case, they read
\begin{equation}\nn
	\left.\frac{1+i{\gamma}^{r}}{2}\psi \right|_{r=R}=0 \,.
\end{equation}
These conditions can also be expressed in terms of the components of the bi-spinor ${\psi}=(\psi_1,\psi_2)^T$ as
\be
	\psi_1(R,\theta)+ie^{-i\theta}\psi_2(R,\theta)=0.
	\label{cc}
\ee

The eigenvalue problem, $H \psi=E\psi$, corresponding to the Hamiltonian in equation (\ref{hamilton}) can be written explicitly as follows
\beq
\left(\begin{array}{cc}
        0 & ie^{-i\theta}({\pa}_r -\frac{i}{r}{\pa}_{\theta})\\
        ie^{i\theta}({\pa}_r +\frac{i}{r}{\pa}_{\theta}) & 0
      \end{array}\right)\left(\begin{array}{c}
                                \psi_1(r,\theta) \\
                               \psi_2(r,\theta)
                              \end{array}\right)= \mathcal{E}\left(\begin{array}{c}
                                \psi_1(r,\theta) \\
                               \psi_2(r,\theta)
                              \end{array}\right)\,,
\label{eigen}\eeq
where $\mathcal{E}= \frac{E}{v_F}$.

Let us first study zero modes, $\mathcal{E}=0$. In this case, one has
\beq \nn
({\pa}_r +\frac{i}{r}{\pa}_{\theta})\psi_1(r,\theta)=0\\
({\pa}_r -\frac{i}{r}{\pa}_{\theta})\psi_2(r,\theta)=0 \nn \,.
\eeq
Thus, the single-valued solutions which are square-integrable at the origin are
\beq \nn
\psi(r,\theta)=\left(\begin{array}{c}
                       \sum_{n=0}^{\infty} a_n r^n e^{in\theta}\\
                       \sum_{l=0}^{\infty} b_l r^l e^{-il\theta}
                     \end{array}
\right)\nn \,,
\eeq
with some arbitrary constants $a_n$, $b_l$. It is easy to check that, when the boundary condition (\ref{cc}) is imposed, no zero mode is left.

In the case $\mathcal{E}\neq 0$, the usual procedure leads to
\beq
\psi(r,\theta)=\sum_{n=-\infty}^{\infty}\left(\begin{array}{c}
                        a_n e^{in\theta}J_n (|\mathcal{E}|r)\\
                     -i \frac{|\mathcal{E}|}{\mathcal{E}}  a_n e^{i(n+1)\theta}J_{n+1} (|\mathcal{E}|r)
                     \end{array}
\right)\,,
\eeq
where $J_n (z)$ is the Bessel function of the first kind of order $n$, and $a_n$ is a normalization factor to be defined in what follows. Note we have already imposed periodicity in $\theta$ and the square integrability at $r=0$. In this case, the boundary condition (\ref{cc}) leads to \cite{BerryM,Ponomarenko2008}
\beq \nn
J_{n-1}(|\mathcal{E}|R)-\frac{|\mathcal{E}|}{\mathcal{E}} J_{n}(|\mathcal{E}|R)=0,\quad n=1,\ldots ,\infty \\
J_{n-1}(|\mathcal{E}|R)+\frac{|\mathcal{E}|}{\mathcal{E}} J_{n}(|\mathcal{E}|R)=0,\quad n=1,\ldots ,\infty \,.
\label{autovalores}\eeq
From these equations, it is clear that the spectrum is symmetric since, changing $\mathcal{E}\rightarrow -\mathcal{E}$ simply interchanges both equations. The allowed positive energy modes are given by
\beq
E^{\mp}_{n,k}=\frac{v_F}{R} {\lambda}^{\mp}_{n-1,k}\,,
\label{ocho}\eeq
where ${\lambda}^{-}_{n-1,k}$ is the $k$-th positive zero of $J_{n-1}(z)- J_{n}(z)$, while ${\lambda}^{+}_{n-1,k}$ is the $k$-th positive zero of $J_{n-1}(z)+ J_{n}(z)$.

The normalized eigenfunctions are thus given by
\beq
\psi_{n,k,\a}^\mp(r,\theta)=a_n e^{in\theta}\left(\begin{array}{c}
                        J_n ( {E}_{n,k}^\mp r/v_F)\\
                     \mp i\a e^{i\theta}J_{n+1} ({E}_{n,k}^\mp r/v_F)
                     \end{array}
\right)\label{autofun}\,,
\eeq
where $\a=\frac{|\mathcal{E}|}{\mathcal{E}}$, and
\beq
	a_n^{-2}=\frac{4 \pi^2 R}{|\mathcal{E}|} J_{n}(|\mathcal{E}|R) J_{n+1}(|\mathcal{E}|R).
\eeq
This model for a graphene disk (also called a circular neutrino billiard) was originally formulated in \cite{BerryM}. In \cite{Ponomarenko2008} it was shown that its predictions are describing relatively well the experimental results for graphene nanodots with diameters $\gtrsim 100$nm.

In the next two sections, we will show that the positive eigenenergies can be ordered in a very precise way and, thus, numbered through a unique index $q,\, q=1,\ldots \infty$, since there is no finite accumulation point for the sequence they form. This being an elliptic boundary problem \cite{Ben-Sant}, the existence of such order is not surprising, but this fact will be useful in finding the gap and in estimating the number of jumps in the mean charge of quantum nanodots for a given range of variation of the gate voltage.


\section{Position of the positive zeroes of $J_{\nu -1}(z)-J_{\nu}(z)$}
\label{sec-dif}

Throughout this and the next sections, $j_{\nu ,k}$ is the $k$-th positive zero of $J_{\nu}(z)$, $j_{\nu ,k}^{\prime}$ is the $k$-th positive zero of $J_{\nu}^{\prime}(z)$, ${\lambda}_{\nu -1 ,k}^{-}$ is the $k$-th positive zero of $J_{\nu-1}(z)-J_{\nu}(z)$, where $\nu \in \mathbb{R}$, $\nu \geq 1$ and $k\in \mathbb{N}\setminus\{0\}$. Most of our results below arise from a detailed consideration of the well-known relations \cite{watson}, valid for $z>0$,
\beq J_{\nu -1}(z)=J_{\nu}^{\prime}(z)+\frac{\nu}{z}J_{\nu}(z)\,,\label{a01}\eeq
\beq J_{\nu}(z)=-J_{\nu-1}^{\prime}(z)+\frac{(\nu-1)}{z}J_{\nu-1}(z)\,.\label{a02}\,,\eeq
and their immediate consequences,
\beq J_{\nu -1}(z)-J_{\nu}(z)=J_{\nu}^{\prime}(z)+\frac{\nu-z}{z}J_{\nu}(z)\,,\label{a1}\eeq
\beq J_{\nu-1}(z)-J_{\nu}(z)=J_{\nu-1}^{\prime}(z)+\frac{z-(\nu-1)}{z}J_{\nu-1}(z)\,.\label{a2}\eeq

We also make use of the equally well-known facts that $j_{\nu ,k}>j_{\nu ,k}^{\prime}>\nu,\,\forall k\geq 1$ and of the interlacing of the zeroes of $J_{\nu}(z)$.

\begin{thm}\label{teorema1}{\hfill\break

\begin{enumerate}

\item \label{item1} ${\lambda}_{\nu -1 ,1}^{-}>\nu$
\item \label{item2} Between two consecutive positive zeroes of $J_{\nu}(z)$, there exists exactly one zero of $J_{\nu-1}(z)-J_{\nu}(z)$ and v.v.
\item \label{item3} Between two consecutive positive zeroes of $J_{\nu-1}(z)$, there exists exactly one zero of $J_{\nu-1}(z)-J_{\nu}(z)$ and v.v.
\item \label{item4} There is no zero of $J_{\nu-1}(z)-J_{\nu}(z)$ in the interval $[j_{\nu-1 ,k},j_{\nu ,k}]$
\item \label{item5} There is exactly one positive zero ${\lambda}_{\nu -1 ,1}^{-}$ between $z=0$ and $z=j_{\nu-1 ,1}$, and it satisfies

\beq j_{\nu-1 ,1}^{\prime}<{\lambda}_{\nu -1 ,1}^{-}<j_{\nu-1 ,1}\label{des01}\eeq
\beq \nu<{\lambda}_{\nu -1 ,1}^{-}<j_{\nu,1}^{\prime}<j_{\nu,1}\label{des02}\eeq

\item \label{item6} $\forall k\geq 1$

\beq j_{\nu-1 ,k}<j_{\nu-1 ,k+1}^{\prime}<{\lambda}_{\nu -1 ,k+1}^{-}<j_{\nu-1 ,k+1}\label{des1}\eeq
\beq j_{\nu ,k}<{\lambda}_{\nu -1 ,k+1}^{-}<j_{\nu,k+1}^{\prime}<j_{\nu,k+1}\label{des2}\eeq

\end{enumerate}

\textbf{Proof of \ref{item1}}:\hfill\break

In a neighborhood of $z=0$
\beq J_{\nu}(z)=\left(\frac{z}{2}\right)^{\nu}\left[\frac{1}{\Gamma (\nu +1)}+\mathcal{O}(z^2)\right]\,. \nn \eeq

As a consequence, for $z\neq 0$ in a neighborhood of $z=0$, $J_{\nu}>0$, $J_{\nu -1}(z)-J_{\nu}(z)>0$ and $J_{\nu}^{\prime}>0$.

Moreover, it is known that $\nu <j_{\nu ,1}^{\prime}<j_{\nu ,1}$ \cite{watson}. Then, both terms on the right hand side of equation (\ref{a1}) are positive $\forall \, 0<z\leq \nu$, and there is no zero of $J_{\nu-1}(z)-J_{\nu}(z)$ in this interval.\hfill\break

\textbf{Proof of \ref{item2}}:\hfill\break

Equation (\ref{a1}) can be rewritten as
\beq
J_{\nu-1}(z)-J_{\nu}(z)=z^{-\nu}e^z\frac{d}{dz}\left(z^{\nu} e^{-z}J_{\nu}(z)\right)\,.
\nn\eeq

Note that between two zeroes of the function $z^{\nu} e^{-z}J_{\nu}(z)$ there is at least one zero of its derivative, and, thus, at least, one zero of $J_{\nu-1}(z)-J_{\nu}(z)$. Now, the zeroes of $z^{\nu} e^{-z}J_{\nu}(z)$ correspond
to zeroes of $J_{\nu}(z)$. Thus, between two consecutive zeroes of $J_{\nu}(z)$ there is, at least, one zero of $J_{\nu-1}(z)-J_{\nu}(z)$.

Moreover, from equations (\ref{a01}) and (\ref{a02}),
\beq
J_{\nu-1}^{\prime}(z)-J_{\nu}^{\prime}(z)&=&J_{\nu}(z)\left(\frac{\nu}{z}-1\right)+J_{\nu-1}(z)\left(\frac{\nu-1}{z}-1\right)\nn \\
&=&J_{\nu}(z)\left(\frac{2\nu}{z}-2-\frac{1}{z}\right)+\left(J_{\nu-1}(z)-J_{\nu}(z)\right)\left(\frac{\nu-1}{z}-1\right)\,.\nn\eeq

Thus,
\beq z^{\nu-1}e^{-z}\frac{d}{dz}\left(z^{1-\nu} e^{z}\left(J_{\nu-1}(z)-J_{\nu}(z)\right)\right)=J_{\nu}(z)\frac{\left(2\nu-2z-1\right)}{z}\,.\eeq
This last equation, together with the fact that ${\lambda}_{\nu -1,1}^{-}>\nu$, imply that between two consecutive zeroes of $J_{\nu-1}(z)-J_{\nu}(z)$ there is, at least, one zero of $J_{\nu}(z)$.

Altogether, there is a unique zero of $J_{\nu-1}(z)-J_{\nu}(z)$ between two consecutive positive zeroes of $J_{\nu}(z)$ and v.v.\hfill\break

\textbf{Proof of \ref{item3}}:\hfill\break

In order to prove that between two consecutive positive zeroes of $J_{\nu-1}(z)$, there exists exactly one zero of $J_{\nu-1}(z)-J_{\nu}(z)$ and v.v. one has to use equation (\ref{a2}), which can be rewritten as
\beq
J_{\nu-1}(z)-J_{\nu}(z)=z^{\nu -1}e^{-z}\frac{d}{dz}\left(z^{1-\nu} e^{z}J_{\nu - 1}(z)\right)\,.
\nn\eeq

Also, from equations (\ref{a01}) and (\ref{a02}),
\beq z^{-\nu}e^{z}\frac{d}{dz}\left(z^{\nu} e^{-z}\left(J_{\nu-1}(z)-J_{\nu}(z)\right)\right)=J_{\nu -1}(z)\frac{\left(2\nu-2z-1\right)}{z}\,.\eeq

The proof is now analog to the latter one. \hfill\break

\textbf{Proof of \ref{item4}}:\hfill\break

In the interval $[j_{\nu-1 ,k},j_{\nu ,k}]$ the functions $J_{\nu-1}(z)$ and $J_{\nu}(z)$ have opposite signs. This, together with the fact that $j_{\nu-1 ,k} \neq j_{\nu ,k}$, imply that $J_{\nu-1}(z)-J_{\nu}(z)$ is either positive or negative, but never zero $\forall z / j_{\nu-1 ,k}\leq z \leq j_{\nu ,k}$.

Combining now what we have found in items \ref{item1}, \ref{item2} and \ref{item3}, it follows that the zero of $J_{\nu-1}(z)-J_{\nu}(z)$ between two consecutive positive zeroes of $J_{\nu}(z)$, and the zero of $J_{\nu-1}(z)-J_{\nu}(z)$ between two consecutive positive zeroes of $J_{\nu -1}(z)$ coincide, and it lies in the interval $(j_{\nu ,k},j_{\nu -1 ,k+1})$.\hfill\break

\textbf{Proof of \ref{item5}}:\hfill\break

As already said, for $z\neq 0$ in a neighborhood of $z=0$, $J_{\nu}>0$, $J_{\nu -1}(z)-J_{\nu}(z)>0$ and $J_{\nu}^{\prime}>0$. From item \ref{item1}, ${\lambda}_{\nu -1 ,1}^{-}>\nu > \nu -1$. Moreover, from equation (\ref{a2}), $J_{\nu -1}(z)-J_{\nu}(z)>0 \, \forall z \in [\nu -1,j_{\nu -1,1}^{\prime}]$. So, $j_{\nu-1 ,1}^{\prime}<{\lambda}_{\nu -1 ,1}^{-}$. Now, evaluating the right hand side of (\ref{a2}) at $z= j_{\nu -1,1}$, it is easy to see that $J_{\nu -1}(j_{\nu -1,1})-J_{\nu}(j_{\nu -1,1})<0$, which shows that there is at least one zero in the interval $j_{\nu-1 ,1}^{\prime}<z<j_{\nu -1,1}$. From item \ref{item3}, there is only one zero of $J_{\nu -1}(z)-J_{\nu}(z)$ in the same interval, and we get (\ref{des01}).

Now, evaluating the right hand side of equation (\ref{a1}) at $j_{\nu ,1}^{\prime}$, we see that ${\lambda}_{\nu -1 ,1}^{-} < j_{\nu ,1}^{\prime}$ and a well-known property \cite{watson} leads to (\ref{des02}).\hfill\break

\textbf{Proof of \ref{item6}}:\hfill\break

Once more, from equation (\ref{a2}), we study the behavior of $J_{\nu-1}(z)-J_{\nu}(z)$, this time between $j_{\nu-1 ,k}$ and $j_{\nu-1 ,k+1}$. For definiteness, we first suppose that $J_{\nu-1}(z)>0$ in the open interval. Then, in a neighborhood of $j_{\nu-1 ,k}$, $J_{\nu-1}^{\prime}(z)>0$. Moreover, $\frac{z-(\nu-1)}{z}>0$, since $k\geq1$ and $j_{\nu-1 ,1}>\nu-1$ \cite{watson}. So, the second term in (\ref{a2}) is always positive. As for the first term, for $j_{\nu-1 ,k}<z<j_{\nu-1 ,k+1}^{\prime}$ it is also positive, since $J_{\nu-1}^{\prime}(z)>0$. As a consequence, $J_{\nu-1}(z)-J_{\nu}(z)$ remains positive till the zero of the derivative is surpassed and, so, there are no zeroes in this interval.

Now, for $j_{\nu-1 ,k+1}^{\prime}<z<j_{\nu-1 ,k+1}$, the second term in (\ref{a2}) is still positive, but the first one is negative and decreases monotonically. Moreover, $J_{\nu-1}(j_{\nu-1 ,k+1})-J_{\nu}(j_{\nu-1 ,k+1})<0$. By item \ref{item3}, there is exactly one root of the difference between $j_{\nu-1 ,k}$ and $j_{\nu-1 ,k+1}$, and it is such that $j_{\nu-1 ,k}<j_{\nu-1 ,k+1}^{\prime}<{\lambda}_{\nu -1 ,k+1}^{-}<j_{\nu-1 ,k+1}$, as stated in equation (\ref{des1}). It is easy to check that the same conclusion is reached by supposing $J_{\nu-1}(z)<0$ in the open interval.

Finally, a similar analysis of the signs of $J_{\nu-1}(z)-J_{\nu}(z)$ from equation (\ref{a1}) leads to the result in (\ref{des2}).

\fin}\end{thm}

\begin{cor}\label{corolario1}{${\lambda}_{\nu,k+1}^{-}<{\lambda}_{\nu-1,k+2}^{-}<{\lambda}_{\nu ,k+2}^{-}$.\hfill\break

\textbf{Proof}: Changing $\nu\rightarrow\nu+1$ in (\ref{des1}), one obtains
\beq \nn j_{\nu ,k}<j_{\nu ,k+1}^{\prime}<{\lambda}_{\nu ,k+1}^{-}<j_{\nu ,k+1}\,.\eeq

Replacing into (\ref{des2}) and, then, using (\ref{des1}) again, this time with $k\rightarrow k+1$,
the interlacing of the roots of $J_{\nu-1}(z)-J_{\nu}(z)$ is proved.\fin } \end{cor}\hfill\break

\section{Position of the positive zeroes of $J_{\nu -1}(z)+J_{\nu}(z)$}
\label{sec-sum}

Throughout this section, $j_{\nu ,k}$ is the $k$-th positive zero of $J_{\nu}(z)$, $j_{\nu ,k}^{\prime}$ is the $k$-th positive zero of $J_{\nu}^{\prime}(z)$, ${\lambda}_{\nu -1 ,k}^{+}$ is the $k$-th positive zero of $J_{\nu-1}(z)+J_{\nu}(z)$, where $\nu \in \mathbb{R}$, $\nu \geq 1$ and $k\in \mathbb{N}\setminus\{0\}$. Most of our results below arise from a detailed consideration of the well-known relations \cite{watson}, valid for $z>0$,

\beq J_{\nu -1}(z)+J_{\nu}(z)=J_{\nu}^{\prime}(z)+\frac{\nu+z}{z}J_{\nu}(z)\,,\label{a3}\eeq
\beq J_{\nu-1}(z)+J_{\nu}(z)=-J_{\nu-1}^{\prime}(z)+\frac{z+(\nu-1)}{z}J_{\nu-1}(z)\,.\label{a4}\eeq

\begin{thm}\label{teorema2}{\hfill\break

\begin{enumerate}

\item \label{itemb1} There is no positive zero of $J_{\nu-1}(z)+J_{\nu}(z)$ in the interval $0< z\leq j_{\nu-1 ,1}$\,.

\item \label{itemb2} Between two consecutive positive zeroes of $J_{\nu}(z)$, there exists exactly one zero of $J_{\nu-1}(z)+J_{\nu}(z)$ and v.v.

\item \label{itemb3} Between two consecutive positive zeroes of $J_{\nu-1}(z)$, there exists exactly one zero of $J_{\nu-1}(z)+J_{\nu}(z)$ and v.v.

\item \label{itemb4} There is exactly one positive zero ${\lambda}_{\nu -1 ,1}^{+}$ between $z=0$ and $z=j_{\nu ,1}$, and it satisfies
\beq j_{\nu,1}^{\prime}<{\lambda}_{\nu -1 ,1}^{+}<j_{\nu,1}\label{des03}\eeq

\item \label{itemb5} There is no zero of $J_{\nu-1}(z)+J_{\nu}(z)$ in the interval $[j_{\nu ,k},j_{\nu-1 ,k+1}]$

\item \label{itemb6} $\forall k\geq 1$

\beq j_{\nu ,k}<j_{\nu ,k+1}^{\prime}<{\lambda}_{\nu -1 ,k+1}^{+}<j_{\nu ,k+1}\label{des3}\eeq
\beq j_{\nu-1 ,k+1}<{\lambda}_{\nu -1 ,k+1}^{+}<j_{\nu-1,k+2}^{\prime}<j_{\nu-1,k+2}\label{des4}\eeq

\end{enumerate}

\hfill\break

\textbf{Proof of \ref{itemb1}}:\hfill\break

In the interval $(0,j_{\nu-1 ,1})$ both $J_{\nu-1}$ and $J_{\nu}$ are positive. Indeed, as already noted at the beginning of Theorem \ref{teorema1}, they are both positive in a neighborhood of the origin, as long as $z\neq 0$; moreover, the first zero of $J_{\nu}$ is bigger than $j_{\nu-1 ,1}$ \cite{watson}. So the statement is proved.\hfill\break

\textbf{Proof of \ref{itemb2}}:\hfill\break

Equation (\ref{a3}) can be rewritten as
\beq
J_{\nu-1}(z)+J_{\nu}(z)=z^{-\nu}e^{-z}\frac{d}{dz}\left(z^{\nu} e^{z}J_{\nu}(z)\right)\,.
\nn\eeq

Thus, between two consecutive positive zeroes of $J_{\nu}(z)$ there is, at least, one zero of $J_{\nu-1}(z)+J_{\nu}(z)$.

Now, from equations (\ref{a01}) and (\ref{a02}),
\beq z^{\nu-1}e^{z}\frac{d}{dz}\left(z^{1-\nu} e^{-z}\left(J_{\nu-1}(z)+J_{\nu}(z)\right)\right)=-J_{\nu}(z)\frac{\left(2\nu+2z-1\right)}{z}\,.\eeq

This last equation, together with the fact that $2\nu+2z-1>0 \, \forall \nu\geq 1$, imply that between two consecutive zeroes of $J_{\nu-1}(z)+J_{\nu}(z)$ there is, at least, one zero of $J_{\nu}(z)$.

Altogether, there is a unique zero of $J_{\nu-1}(z)+J_{\nu}(z)$ between two consecutive positive zeroes of $J_{\nu}(z)$ and v.v.\hfill\break

\textbf{Proof of \ref{itemb3}}:\hfill\break

In order to prove that between two consecutive positive zeroes of $J_{\nu-1}(z)$, there exists exactly one zero of $J_{\nu-1}(z)+J_{\nu}(z)$ and v.v. one has to use equation (\ref{a4}), which can be rewritten as
\beq
J_{\nu-1}(z)+J_{\nu}(z)=-z^{\nu -1}e^{z}\frac{d}{dz}\left(z^{1-\nu} e^{-z}J_{\nu - 1}(z)\right)\,.
\nn\eeq

Also, from equations (\ref{a01}) and (\ref{a02}),
\beq z^{-\nu}e^{-z}\frac{d}{dz}\left(z^{\nu} e^{z}\left(J_{\nu-1}(z)+J_{\nu}(z)\right)\right)=J_{\nu -1}(z)\frac{\left(2\nu+2z-1\right)}{z}\,.\eeq

The proof is now analog to the latter one. \hfill\break

\textbf{Proof of \ref{itemb4}}:\hfill\break

As already said, for $z\neq 0$ in a neighborhood of $z=0$, $J_{\nu}>0$, $J_{\nu -1}(z)+J_{\nu}(z)>0$ and $J_{\nu}^{\prime}>0$. From the r.h.s. of equation (\ref{a3}), $J_{\nu -1}(z)+J_{\nu}(z)>0\, \forall z\in (0,j_{\nu ,1}^{\prime}]$.

 Now, evaluating the right hand side of (\ref{a3}) at $z= j_{\nu,1}$, it is easy to see that $J_{\nu -1}(j_{\nu,1})+J_{\nu}(j_{\nu,1})<0$, which shows that there is at least one zero in the interval $j_{\nu ,1}^{\prime}<z<j_{\nu,1}$. From item \ref{itemb2}, there is only one zero of $J_{\nu -1}(z)+J_{\nu}(z)$ in the same interval, and we get (\ref{des03}).\hfill\break

 \textbf{Proof of \ref{itemb5}}:\hfill\break

In the interval $(j_{\nu ,k},j_{\nu-1 ,k+1})$ the functions $J_{\nu-1}(z)$ and $J_{\nu}(z)$ have the same sign. This, together with the fact that $j_{\nu-1 ,k+1} \neq j_{\nu ,k}$, imply that $J_{\nu-1}(z)+J_{\nu}(z)$ is either positive or negative, but never zero $\forall z / j_{\nu ,k}\leq z \leq j_{\nu-1 ,k+1}$.

Combining now what we have found in items \ref{itemb1}, \ref{itemb2}, \ref{itemb3} and \ref{itemb4}, it follows that the zero of $J_{\nu-1}(z)+J_{\nu}(z)$ between $j_{\nu-1 ,k+1}$ and $j_{\nu-1 ,k+2}$ (${\lambda}_{\nu -1 ,k+1}^{+}$) is the same as the zero of $J_{\nu-1}(z)+J_{\nu}(z)$ between $j_{\nu,k}$ and $j_{\nu ,k+1}$, and it lies in the interval $(j_{\nu-1 ,k+1},j_{\nu,k+1})$.\hfill\break

\textbf{Proof of \ref{itemb6}}:\hfill\break

Once more, from equation (\ref{a4}), we study the behavior of $J_{\nu-1}(z)+J_{\nu}(z)$, this time between $j_{\nu-1 ,k+1}$ and $j_{\nu-1 ,k+2}$. For definiteness, we first suppose that $J_{\nu-1}(z)>0$ in the open interval. Then, in a neighborhood of $j_{\nu-1 ,k+1}$, $J_{\nu-1}^{\prime}(z)>0$. Moreover, $\frac{z+(\nu-1)}{z}>0$. So, the second term in the r.h.s of (\ref{a4}) is always positive. As for the first term, for $j_{\nu-1 ,k+2}^{\prime}<z<j_{\nu-1 ,k+2}$ it is also positive, since $J_{\nu-1}^{\prime}(z)<0$. As a consequence, $J_{\nu-1}(z)+J_{\nu}(z)$ remains positive in this interval and, so, there are no zeroes here.

Now, for $j_{\nu-1 ,k+1}<z<j_{\nu-1 ,k+2}^{\prime}$, the second term in the r.h.s. of (\ref{a4}) is still positive, but the first one is negative . Moreover, $J_{\nu-1}(j_{\nu-1 ,k+1})+J_{\nu}(j_{\nu-1 ,k+1})<0$. By item \ref{itemb3}, there is exactly one root of the difference between $j_{\nu-1 ,k+1}$ and $j_{\nu-1 ,k+2}$, and it is such that $j_{\nu-1 ,k+1}<{\lambda}_{\nu -1 ,k+1}^{+}<j_{\nu-1 ,k+2}^{\prime}<j_{\nu-1 ,k+2}$, as stated in equation (\ref{des4}). It is easy to check that the same conclusion is reached by supposing $J_{\nu-1}(z)<0$ in the open interval.

Finally, a similar analysis of the signs of $J_{\nu-1}(z)+J_{\nu}(z)$ from equation (\ref{a3}) leads to the result in (\ref{des3}).

\fin } \end{thm}

\begin{cor}{\beq {\lambda}_{\nu -1 ,k+1}^{+}<{\lambda}_{\nu -2 ,k+2}^{+}<{\lambda}_{\nu -1 ,k+2}^{+}\,.\eeq \hfill\break

\textbf{Proof}:\hfill\break

This follows immediately, by combining (\ref{des3}) with (\ref{des4}) and iterating.\fin

}\end{cor}\hfill\break

\begin{cor}{\beq j_{\nu ,k}<{\lambda}_{\nu -1 ,k+1}^{-}<j_{\nu ,k+1}^{\prime}<{\lambda}_{\nu -1 ,k+1}^{+}<j_{\nu ,k+1}\label{des5}\eeq \hfill\break

\textbf{Proof}:\hfill\break

This follows immediately, by combining (\ref{des2}) with (\ref{des3}).\fin

}\end{cor}\hfill\break

\begin{cor}{Between two consecutive zeros of $J_{\nu-1}(z)-J_{\nu}(z)$ there is one zero of $J_{\nu-1}(z)+J_{\nu}(z)$ and v.v.\hfill\break

\textbf{Proof}:\hfill\break

Immediate, by iterating (\ref{des5}).\fin

}\end{cor}\hfill\break

\begin{cor}{\beq j_{\nu-1 ,k}<{\lambda}_{\nu -1 ,k}^{+}<j_{\nu-1 ,k+1}^{\prime}<{\lambda}_{\nu -1 ,k+1}^{-}<j_{\nu-1 ,k+1}\label{des6}\eeq \hfill\break

\textbf{Proof}:\hfill\break

This follows immediately, by combining (\ref{des1}) with (\ref{des4}).\fin

}\end{cor}\hfill\break

\begin{cor}{The zeroes of the derivatives of Bessel functions of the first kind are interlaced according to

\beq \nn
j_{\nu ,k+1}^{\prime}<j_{\nu-1 ,k+2}^{\prime}<j_{\nu ,k+2}^{\prime}
\eeq
\hfill\break

\textbf{Proof}:\hfill\break

Immediate, combining (\ref{des5}) and (\ref{des6}).\fin

}\end{cor}\hfill\break

\section{Mean charge}\label{sec-charge}

In order to obtain the zero temperature mean charge in the presence of a gate potential (chemical potential) $\mu$ in a field-theoretical approach, one needs the Green function, $G$, of the Dirac operator in equation (\ref{diracop}), with its temporal derivative shifted by $-i\mu$. To this end, as in \cite{bfsv}, we will first determine the Green function $\mathcal{G}$ of the auxiliary operator $\mathcal{D}=\gamma^0\slashed{D}$. Now, it is easy to show that $G=\mathcal{G}{\gamma}^0$. In turn, $\mathcal{G}$ is given by
\begin{equation}
\mathcal{G}(x,x^{\prime})=\sum_{q,\alpha}\int dk_0
    \frac{\psi_{q}(\vec{x})e^{-ik_0 x^0}\otimes\psi_{q}^\dag (\vec{x}^{\prime})e^{ik_0 x^{\prime 0}}}
    {k_0 + \mu +{\alpha}{E_q}}\,, \label{g}
\end{equation}
where $\alpha=\pm 1$ is defined as in Section \ref{sec-conventions}, $\psi_{q}(\vec{x})$ are the eigenfunctions (\ref{autofun}) corresponding to $|\mathcal{E}| =E_q/v_F $, and $q$ labels the order in which the eigenvalues in
equation (\ref{ocho}) appear as the absolute value of the energy grows.

As is well-known \cite{chodos}, we can express the local mean number of charge carriers, $n(\mu ,x)$, as
\begin{equation}
    n (\mu ,x) = -i \, {\rm tr} (\gamma^0 G(x,x))=-i \, {\rm tr} (\mathcal{G}(x,x))\,.\label{j01}
\end{equation}
By using (\ref{autofun}) and (\ref{g}) we obtain from (\ref{j01}) the mean number of charge carriers in the quantum nanodot as
\beq
n(\mu)&\equiv&
	\int_0^R r dr \int_0^{2\pi} d\theta \,  n(\mu ,x) =
	-i \sum_{\alpha,q}\int_{-\infty}^{\infty} \frac{dk_0}{2\pi} \frac {1}{k_0 + \mu +{\alpha}{E_q}} \nn \\
&=& -i \sum_{q}\int_{-\infty}^{\infty} \frac{dk_0}{2\pi} \frac {2(k_0 + \mu)}{(k_0 + \mu)^2 - {E_q}^2}
\,.
\label{j02}
\eeq
This last expression obviously needs an ultraviolet regularization. As in \cite{bfsv}, we will follow the lines of \cite{chodos}. In the first place we shift, in the denominator, $k_0 \rightarrow k_0+i\epsilon \, {\rm sgn}(k_0)$, which corresponds to the Feynman prescription for treating the poles on the real axes in the integral (\ref{j02}). Thus, we get

\beq
n(\mu)=-i \sum_{q}\int_{-\infty}^{\infty} \frac{dk_0}{2\pi} \frac {2(k_0  + \mu)}{(k_0 + i\epsilon \, {\rm sgn}(k_0) + \mu)^2 - {E_q}^2}
\,. \nn
\eeq

Now, we change the integration variable to $t=k_0 + \mu$ to get

\beq
n(\mu)=-i \sum_{q}\int_{-\infty}^{\infty} \frac{dt}{2\pi} \frac {2t }{(t + i\epsilon \, {\rm sgn}(t- \mu))^2 - {E_q}^2}
\,, \nn
\eeq
or, equivalently,

\beq
n(\mu)=-\frac{i}{2} \sum_{q}\left\{\int_{-\infty}^{\infty} \frac{dt}{2\pi} \frac {2t }{(t + i\epsilon \, {\rm sgn}(t- \mu))^2 - {E_q}^2} -
\int_{-\infty}^{\infty} \frac{dt}{2\pi} \frac {2t }{(t + i\epsilon \, {\rm sgn}(t+ \mu))^2 - {E_q}^2}\right\}
\,, \nn
\eeq
which leads to

\beq
n(\mu)&=&\frac{1}{\pi} \sum_{q}\int_{-\infty}^{\infty} dt \frac {t^2 \epsilon \left [{\rm sgn}(t+ \mu)-{\rm sgn}(t- \mu)\right ] }{\left[(t+ i\epsilon \, {\rm sgn}(t- \mu))^2 - {E_q}^2\right]\left[(t+ i\epsilon \, {\rm sgn}(t+ \mu))^2 - {E_q}^2\right]}
 \nn \\
&=& \frac{2}{\pi}\sum_{q}\int_{-|\mu|}^{|\mu|} dt \frac {\epsilon \, t^2}{\left[(t+ i\epsilon \, {\rm sgn}(t- \mu))^2 - {E_q}^2\right]\left[(t+ i\epsilon \, {\rm sgn}(t+ \mu))^2 - {E_q}^2\right]}\,.
\nn\eeq
Now, rewriting the integrand in terms of simple fractions and using that, for $\epsilon \rightarrow 0^+$,\hfill\break $\frac{1}{x\pm i\epsilon}= PV\left(\frac{1}{x}\right)\mp i\pi \delta(x)$, we obtain, for one valley and one spin value,
\begin{equation}
   n(\mu)={\rm sgn} (\mu) \sum_{q}\Theta(|\mu|-|E_q|) \,.\label{j04}
\end{equation}
This result was to be expected, since the manifold we are studying is compact, and so, the spectrum (equation (\ref{ocho})) is discrete. Equation (\ref{j04}) thus confirms that for electrons (holes), every filled state contributes with 1 ($-1$) to the mean number.

Though the actual determination of the precise values of the gate potential at which the jumps occur for a graphene disk requires a numerical evaluation, the mathematical results obtained in the two previous sections put a bound on the number of equations to be considered in the system (\ref{autovalores}) once a given range of $\mu$ is selected.

Note that our results in the two previous sections also guarantee that, for a finite range of variation of the gate potential, $n(\mu)$ will show a finite number of jumps, all of them of the same height, since the spectrum is non-degenerate. In fact, this follows immediately from the interlacing with the zeroes of the Bessel functions, whose multiplicity is one.

Here, we remind the reader 
that the mean charge of a given nanodot is the product of $n(\mu)$ with the charge of the electron.
On the right-hand side of equation (\ref{j04}) we recognize the counting function for positive eigenvalues, $N(\mu) = \sum_{q}\Theta(|\mu|-|E_q|)$. Applying the results of \cite{BerryM}, one can show that its asymptotics for a smoothed staircase is given  by
\be
	\<N(\mu) \> \mathop\simeq\frac{A (\mu/ v_F)^2}{4\pi}+\ldots 
\label{N}
\ee
where $A=\pi R^2$ is the area of the quantum dot.  Note that this smoothening is the result of a semiclasssical approximation with a continous spectrum at zero temperature, see  \cite{BerryM},\cite{BaltesHilf}.

\begin{figure}[here]
\centering \includegraphics[width=10cm]{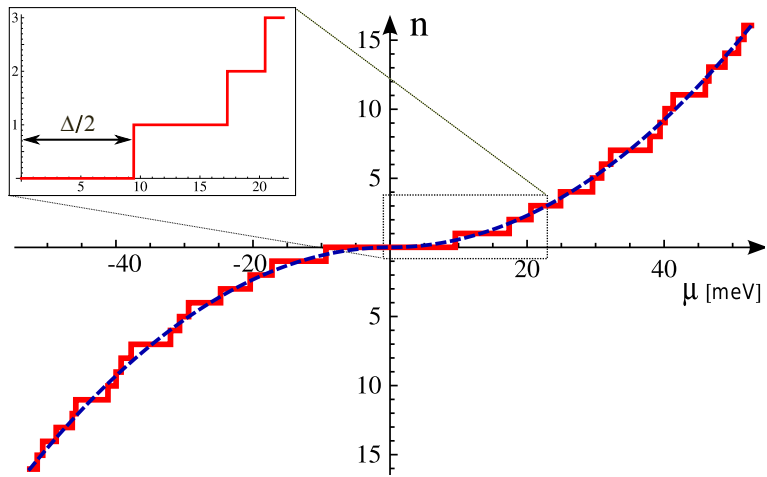}
\caption{
The number of charge carriers (solid red line) for a single fermion species as a function of the chemical potential $\mu$[meV] (positive/negative values correspond to the number of electrons/holes) for a dot of radius $R=100$nm. The blue dashed line is the smoothed expression $\<N\>$ in equation (\ref{N}).
\label{fig}}
\end{figure}

The behavior of the number of charge carriers for a quantum dot with radius $R=100$nm, as a function of $\mu $, is shown in Figure \ref{fig} (red solid line) along with the smoothed approximate expression (blue dashed line) given by (\ref{N}). Note that our results in sections \ref{sec-dif} and \ref{sec-sum} guarantee that no other jumps occur  within the interval considered. The results in the same sections also show that the gap $\Delta$ is given by
$$
	\Delta= 2\frac{\hbar c v_F}{R} {\lambda}^{-}_{0,1}
	\approx 2.8 \frac{\hbar c v_F}{R}.
$$
which for $R=100$nm gives $\Delta\approx 18.4$meV. We reestablished here $\hbar$ and $c$ for a better comparison with experimental results. This result is in qualitative agreement with both experimental and theoretical research \cite{Schnez2008,Schnez2009,Ponomarenko2008}.

The distances between consecutive eigenvalues $E_q$, (equivalently, the length of the steps in $n(\mu$)) are related to the width of the Coulomb diamonds in transport measurements performed on nanodots. We have checked that the  statistical distribution for these distances, calculated from the first $2496$ energy levels, nicely fits a Poissonian distribution \cite{BerryM}. As shown in  \cite{Ponomarenko2008}, such distribution also fits the experimental results for nanodots of intermediate size. Unfortunately, current experiments have not yet performed a comparison of the steps themselves, which have a unique sequencing pattern along the whole range of variation of $\mu$. Such comparison would be crucial for determining the viability of this model as a description of graphene quantum dots, and would also help explaining some experimental features, which depend on the actual number of charge carriers in a quantum dot \cite{Schnez2008}.

The finite-temperature mean number can be obtained by performing a Wick rotation to Euclidean space-time and replacing the integral over $k_0$ in equation (\ref{j02}) by a sum over Matsubara frequencies $\frac{(2l+1)\pi}{\beta}$, where (again in natural units) $\beta=\frac{1}{T}$, with $T$ being the temperature. The result of a direct calculation is
\begin{equation}
   n(\mu)={\rm sgn} (\mu) \sum_{q}\left[\left(1+e^{(|E_q|-|\mu|)\beta}\right)^{-1} - \left(1+e^{(|E_q|+|\mu|)\beta}\right)^{-1}\right] \,.
\end{equation}

This result is the one expected from the Fermi-Dirac distribution. It reduces to our expression in equation (\ref{j04}) in the zero-temperature ($\beta \rightarrow \infty$) limit. As the temperature grows, the steps become increasingly smoother.

\section{Some comments and remarks}\label{sec-remarks}

From a mathematical point of view, the main results of this article are contained in theorems \ref{teorema1} and \ref{teorema2} and their corollaries. Altogether, we have established the precise positions of zeroes of sums and differences of Bessel functions of the first kind of orders differing by one, for real nonnegative orders, as well as their interlacing among themselves and with the zeroes of Bessel functions of the same kind and of their derivatives. As a byproduct we also got a new proof of the interlacing of the zeroes of the derivatives of Bessel functions of the first kind. Other proofs of this property were presented quite recently in \cite{zou,palmai,palmai2}. Our physical problem involves only Bessel functions of integer orders; however, our results are more general and hold for all real non--negative orders. Note that some of these results could also have been obtained, in a different approach, by making use of the results given in Lemma 4 of reference \cite{palmai2}. For instance, once item \ref{item2} in Theorem \ref{teorema1} has been proved, the result in item \ref{item3} of the same theorem can be shown to hold from the transitivity property in such lemma.

As a physical motivation for our mathematical problem, we presented the spectrum of graphene nanodots with the geometry of a disk, and calculated their mean charge and its dependence on the external gate voltage (chemical potential). Such dependence was determined in Section \ref{sec-charge}, and made visible in figure \ref{fig}. Though the actual determination of the precise values of the gate potential at which such jumps occur requires a numerical evaluation, the mathematical results detailed in the previous paragraph put a bound on the number of equations to be considered in the system (\ref{autovalores}) once a given range of $\mu$ is selected. Note that, as is to be expected for a finite sample, the quantum conductance of a graphene disk ($e^2\frac{\partial n}{\partial \mu}$) presents successive peaks, appearing each time the gate voltage goes through an allowed energy value. Such peaks are not equally spaced for our geometry and boundary conditions. The experimental study of graphene nanodots is still at a very preliminary stage and concerns mainly transport properties \cite{KatsBook}, but some indications in favour of the model considered here are already evident \cite{Ponomarenko2008}.

In the case of Berry--Mondragon boundary conditions one thing is for sure: no radial current is allowed to enter or leave the dot. The same is true of any local boundary condition leading to a self adjoint Hamiltonian, the zig-zag boundary condition among others \cite{Ben-Sant}. However, the existence of a tangential current, forbidden in the zig-zag case, is not forbidden by Berry-Mondragon boundary conditions. Its properties and relevance for electronic devices constitute a subject worth exploring.

Finally, it is worth mentioning that, as noted in reference \cite{tesis}, our theorems are also useful when studying field theories in higher-dimensional bounded regions with cylindrical symmetry.


\section*{Acknowledgements}
This work was supported in part by FAPESP (I.V.F.). Work of C.G.B. and E.M.S. was supported by UNLP (Proyecto 11/X615), CONICET (PIP0681) and ANPCyT (PICT0605).


\end{document}